\documentclass[lettersize,journal]{IEEEtran}
\pdfoutput=1
\usepackage{amsmath,amsfonts}
\usepackage{algorithmic}
\usepackage{algorithm}
\usepackage{array}
\usepackage[caption=false,font=normalsize,labelfont=sf,textfont=sf]{subfig}
\usepackage{textcomp}
\usepackage{stfloats}
\usepackage{url}
\usepackage{verbatim}
\usepackage{graphicx}
\graphicspath{{./}}
\usepackage{cite}
\usepackage{tikz}
\usetikzlibrary{decorations.pathreplacing, calc, arrows.meta}
\makeatletter
\tikzset{
    node distance=0.4cm,%
    database/.style={
        path picture={
            \draw (0, 1.5*\database@segmentheight) circle [x radius=\database@radius,y radius=\database@aspectratio*\database@radius];
            \draw (-\database@radius, 0.5*\database@segmentheight) arc [start angle=180,end angle=360,x radius=\database@radius, y radius=\database@aspectratio*\database@radius];
            \draw (-\database@radius,-0.5*\database@segmentheight) arc [start angle=180,end angle=360,x radius=\database@radius, y radius=\database@aspectratio*\database@radius];
            \draw (-\database@radius,1.5*\database@segmentheight) -- ++(0,-3*\database@segmentheight) arc [start angle=180,end angle=360,x radius=\database@radius, y radius=\database@aspectratio*\database@radius] -- ++(0,3*\database@segmentheight);
        },
        minimum width=2*\database@radius + \pgflinewidth,
        minimum height=3*\database@segmentheight + 2*\database@aspectratio*\database@radius + \pgflinewidth,
    },
    database segment height/.store in=\database@segmentheight,
    database radius/.store in=\database@radius,
    database aspect ratio/.store in=\database@aspectratio,
    database segment height=0.1cm,
    database radius=0.25cm,
    database aspect ratio=0.35,
    basestation/.style={
        path picture={
            \draw \foreach \X in {\basestation@towerwidth/6,\basestation@towerwidth/2} {(-\X,\basestation@towerheight/6) |- (\X,0) -- (\X,\basestation@towerheight/6)} (-\basestation@towerwidth/2, -\basestation@towerheight) -- (-\basestation@towerwidth/12, 0) coordinate[pos=0.175] (l3)
                coordinate[pos=0.5] (l2) coordinate[pos=0.775] (l1)
                (\basestation@towerwidth/2, -\basestation@towerheight) -- (\basestation@towerwidth/12, 0) coordinate[pos=0.175] (r3)
                coordinate[pos=0.5] (r2) coordinate[pos=0.775] (r1);
            \draw (l1) -- (r1) (l2) -- (r2) (l3) -- (r3)
                (l1) -- (r2) (l2) -- (r3) (l3) -- (r2) (l2) -- (r1);
            },
            minimum height=\basestation@towerheight + 3.0cm + \pgflinewidth,
            minimum width=\basestation@towerwidth + \pgflinewidth,
        },
        basestation tower height/.store in=\basestation@towerheight,
        basestation tower width/.store in=\basestation@towerwidth,
        basestation tower height=3cm,
        basestation tower width=\basestation@towerheight/2,
    mybox/.style={rectangle, draw=black, line width=.30mm, inner sep=3pt, minimum width=1.6cm, text width=1.55cm, align=center},%
    mymaneuverarrow/.style={-stealth, shorten >=1pt, line width=0.33mm, thick},
    mycommunicationarrow/.style={->, shorten >=1pt, line width=0.7mm, ultra thick},
    mycommunicationarrowdouble/.style={<->, shorten >=1pt, line width=0.7mm, ultra thick},
    radiation/.style={{decorate, decoration={expanding waves, angle=45, segment length=0.1cm}}},
}
\makeatother

\usetikzlibrary{svg.path}
\usepackage{scalerel}
\definecolor{orcidlogocol}{HTML}{A6CE39}
\makeatletter
\tikzset{
    orcidlogo/.pic={
        \fill[orcidlogocol] svg{M256,128c0,70.7-57.3,128-128,128C57.3,256,0,198.7,0,128C0,57.3,57.3,0,128,0C198.7,0,256,57.3,256,128z};
        \fill[white] svg{M86.3,186.2H70.9V79.1h15.4v48.4V186.2z}
                     svg{M108.9,79.1h41.6c39.6,0,57,28.3,57,53.6c0,27.5-21.5,53.6-56.8,53.6h-41.8V79.1z M124.3,172.4h24.5c34.9,0,42.9-26.5,42.9-39.7c0-21.5-13.7-39.7-43.7-39.7h-23.7V172.4z}
                     svg{M88.7,56.8c0,5.5-4.5,10.1-10.1,10.1c-5.6,0-10.1-4.6-10.1-10.1c0-5.6,4.5-10.1,10.1-10.1C84.2,46.7,88.7,51.3,88.7,56.8z};
    }
}
\makeatother
\newcommand\orcidicon[1]{\textsuperscript{\href{https://orcid.org/#1}{\mbox{\scalerel*{
\begin{tikzpicture}[yscale=-1,transform shape]
\pic{orcidlogo};
\end{tikzpicture}
}{|}}}}}

\usepackage[colorinlistoftodos]{todonotes}
\usepackage{color,soul}
\usepackage{xparse}

\usepackage{siunitx}

\usepackage[nolist]{acronym}
\begin{acronym}
    \acro{3gpp}[3GPP]{Third Generation Partnership Project}
    \acro{5g}[5G]{Fifth Generation Cellular Communication}
    \acro{5gaa}[5GAA]{5G Automotive Association}
    \acro{acea}[ACEA]{European Automobile Manufacturers Association}
    \acro{ad}[AD]{automated driving}
    \acroindefinite{ad}{an}{an}
    \acro{adas}[ADAS]{advanced driver assistance system}
    \acroindefinite{adas}{an}{an}
    \acro{ai}[AI]{artificial intelligence}
    \acro{ap}[AP]{access point}
    \acroindefinite{ap}{an}{an}
    \acro{arq}[ARQ]{automatic repeat request}
    \acroindefinite{arq}{an}{an}
    \acro{as}[AS]{autonomous system}
    \acroindefinite{as}{an}{an}
    \acro{avp}[AVP]{automated valet parking}
    \acroindefinite{avp}{an}{an}
    \acro{bs}[BS]{base station}
    \acro{bsm}[BSM]{Basic Safety Message}
    \acro{c2ccc}[C2C-CC]{Car-to-Car Communication Consortium}
    \acro{ca}[CA]{certificate authority}
    \acroplural{ca}[CAs]{Certificate Authorities}
    \acro{cacc}[CACC]{cooperative adaptive cruise control}
    \acro{cads}[CADS]{cooperative automated driving systems}
    \acro{cam}[CAM]{Cooperative Awareness Message}
    \acro{camp}[CAMP]{Crash Avoidance Metrics Partnership}
    \acro{cav}[CAV]{connected and automated vehicle}
    \acro{catarc}[CATARC]{China Automotive Technology and Research Center Co, Ltd.}
    \acro{cbr}[CBR]{channel busy ratio}
    \acro{ccam}[CCAM]{cooperative, connected, and automated mobility}
    \acro{ccsa}[CCSA]{China Communication Standards Association}
    \acro{cdas}[CDAS]{cooperative driver assistance system}
    \acro{ciss}[CISS]{cooperative integrated safety system}
    \acro{cits}[C-ITS]{cooperative intelligent transport system}
    \acro{clc}[CLC]{cooperative lane change}
    \acro{clcs}[CLCS]{cooperative lane change service}
    \acro{cmc}[CMC]{Connected Motorcycle Consortium}
    \acro{cmp}[CMP]{collaborative maneuver protocol}
    \acro{cnn}[CNN]{convolutional neural networks}
    \acro{cpm}[CPM]{Cooperative Perception Message}
    \acro{cqm}[CQM]{Cooperative Request Message}
    \acro{crm}[CRM]{Cooperative Response Message}
    \acro{csm}[CSM]{Cooperative Sensing Message}
    \acro{csti}[CSTI]{Council for Science, Technology and Innovation}
    \acro{cpu}[CPU]{central processing unit}
    \acro{cv2x}[C-V2X]{cellular V2X}
    \acro{cvip}[CVIP]{Complex Vehicular Interactions Protocol}
    \acro{d2d}[D2D]{device-to-device}
    \acro{de}[DE]{data element}
    \acro{denm}[DENM]{Dedicated Environment Notification Message}
    \acro{df}[DF]{data field}
    \acro{dsrc}[DSRC]{dedicated short-range communication}
    \acro{dtn}[DTN]{delay-tolerant network}
    \acro{ec}[EC]{European Commission}
    \acro{ecu}[ECU]{electronic control unit}
    \acroindefinite{ecu}{an}{an}
    \acro{eebl}[EEBL]{electronic emergency break light}
    \acroindefinite{eebl}{an}{an}
    \acro{enb}[eNB]{evolved NodeB}
    \acroindefinite{enb}{an}{an}
    \acro{epm}[EPM]{Environmental Perception Message}
    \acroindefinite{epm}{an}{an}
    \acro{etsi}[ETSI]{European Telecommunications Standards Institute}%
    \acro{fcc}[FCC]{Federal Communcations Commission}
    \acro{fhwa}[FHWA]{Federal Highway Administration}%
    \acro{gdpr}[GDPR]{General Data Protection Regulation}
    \acro{hv}[HV]{host vehicle}
    \acroindefinite{hv}{an}{a}
    \acro{i2v}[I2V]{infrastructure-to-vehicle}
    \acroindefinite{i2v}{an}{an}
    \acro{icn}[ICN]{information-centric network}
    \acroindefinite{icn}{an}{an}
    \acro{icv}[ICV]{intelligent and connected vehicle}
    \acroindefinite{icv}{an}{an}
    \acro{ieee}[IEEE]{Institute of Electrical and Electronics Engineers}%
    \acro{iot}[IoT]{Internet of things}
    \acro{iov}[IoV]{Internet of vehicles}
    \acroindefinite{iov}{an}{an}
    \acro{iso}[ISO]{International Organization for Standardization}
    \acroindefinite{iov}{an}{an}
    \acro{its}[ITS]{intelligent transport system}
    \acroindefinite{its}{an}{an}
    \acro{itu}[ITU]{International Telecommunication Union}
    \acro{kpi}[KPI]{key performance indicator}
    \acro{los}[LoS]{line-of-sight}
    \acro{lte}[LTE]{Long-Term Evolution}
    \acroindefinite{lte}{an}{a}
    \acro{ltev}[LTE-V]{vehicular \acs{lte}}
    \acroindefinite{ltev}{an}{a}
    \acro{mac}[MAC]{medium access control}
    \acro{mcm}[MCM]{Maneuver Coordination Message}
    \acroindefinite{mcm}{an}{a}
    \acro{mcs}[MCS]{Maneuver Coordination Service}
    \acroindefinite{mcm}{an}{a}
    \acro{manet}[MANET]{mobile ad-hoc network}
    \acro{mec}[MEC]{multi-access edge computing}
    \acro{miit}[MIIT]{Ministry of Industry and Information Technology}
    \acro{mfm}[MFM]{Maneuver Feedback Message}
    \acroindefinite{mfm}{an}{a}
    \acro{molit}[MOLIT]{Ministry of Land, Infrastructure and Transport}
    \acro{mptcp}[MPTCP]{Multipath Transmission Control Protocol}
    \acroindefinite{mptcp}{an}{a}
    \acro{mpc}[MPC]{model predictive control}
    \acroindefinite{mpc}{an}{a}
    \acro{msm}[MSM]{Maneuver Status Message}
    \acroindefinite{msm}{an}{a}
    \acro{naas}[NaaS]{network as a service}
    \acro{ndrc}[NDRC]{National Development and Reform Commission}%
    \acro{ndn}[NDN]{named data networking}%
    \acro{ngmn}[NGMN]{Next Generation Mobile Networks}%
    \acro{nprm}[NPRM]{Notice of Proposed Rulemaking}
    \acroindefinite{nprm}{an}{a}
    \acro{oem}[OEM]{original equipment manufacturer}
    \acroindefinite{oem}{an}{an}
    \acro{ota}[OTA]{over-the-air}
    \acroindefinite{ota}{an}{an}
    \acro{pdr}[PDR]{packet delivery ratio}
    \acro{per}[PER]{packet error rate}
    \acro{pki}[PKI]{public key infrastructure}
    \acro{plr}[PLR]{packet loss ratio}
    \acro{prr}[PRR]{packet reception ratio}
    \acro{qoe}[QoE]{quality of experience}
    \acro{qos}[QoS]{quality of service}
    \acro{rat}[RAT]{radio access technology}
    \acro{rto}[RTO]{retransmission timeout}
    \acroindefinite{rto}{an}{a}
    \acro{rsu}[RSU]{road side unit}
    \acroindefinite{rsu}{an}{a}
    \acro{rse}[RSE]{road side equipment}
    \acroindefinite{rse}{an}{a}
    \acro{rv}[RV]{remote vehicle}
    \acroindefinite{rv}{an}{a}
    \acro{sae}[SAE]{SAE International}
    \acro{sc}[SC]{steering committee}
    \acro{scms}[SCMS]{security credential management system}
    \acro{sdn}[SDN]{software-defined network}
    \acro{sdo}[SDO]{standards developing organization}
    \acro{sdr}[SDR]{software-defined radio}
    \acro{slr}[SLR]{service-level requirement}
    \acro{sotif}[SOTIF]{safety of the intended functionality}
    \acro{sps}[SPS]{semi-persistent scheduling}
    \acro{tc}[TC]{technical committee}
    \acro{tcp}[TCP]{Transmission Control Protocol}
    \acro{tdm}[TDM]{time-division multiplexing}
    \acro{tme}[TME]{traffic management entity}
    \acro{tms}[TMS]{traffic management system}
    \acro{uav}[UAV]{unmanned aerial vehicle}
    \acro{ue}[UE]{user equipment}
    \acro{us}[US]{United States}
    \acro{v2i}[V2I]{vehicle-to-infrastructure}
    \acro{v2n}[V2N]{vehicle-to-network}
    \acro{v2p}[V2P]{vehicle-to-pedestrian}
    \acro{v2r}[V2R]{vehicle-to-roadside Infrastructure}
    \acro{v2v}[V2V]{vehicle-to-vehicle}
    \acro{v2x}[V2X]{vehicle-to-everything}
    \acro{vanet}[VANET]{vehicular ad-hoc network}
    \acro{vda}[VDA]{Verband der Automobilindustrie}
    \acro{vlc}[VLC]{visible light communication}
    \acro{vru}[VRU]{vulnerable road user}
    \acro{vsn}[VSN]{vehicular social network}
    \acro{wg}[WG]{working group}
    \acro{wlan}[WLAN]{Wireless Local Area Network}
    \acro{wsn}[WSN]{wireless sensor network}
\end{acronym}

\acrodefplural{ca}[CAs]{Certificate Authorities}
\acrodefplural{as}[ASes]{Autonomous Systems}

\usepackage{url}
\usepackage{hyperref}
\hypersetup{
    colorlinks,
    allcolors={black}
}

\usepackage[nameinlink,capitalize]{cleveref} % For \cref instead of \autoref, for correct equation cross-referencing

\usepackage{xspace}
\makeatletter
\let\@autoref=\autoref
\renewcommand*{\autoref}[2][]{\ifthenelse{\equal{#1}{}}{\@autoref{#2}}{\hyperref[#1]{\begin{NoHyper}\@autoref{#2}~\subref{#1}\end{NoHyper}}}\xspace}
\makeatother

\begin{document}

\title{Vehicular Cooperative Maneuvers -- Quo Vaditis?}

\ifCLASSOPTIONpeerreview
    \author{Anonymous Authors}
\else
\author{Bernhard~H\"afner\,\orcidicon{0000-0002-1383-1052}, J\"org~Ott\,\orcidicon{0000-0001-8311-8036}, and Georg Albrecht Schmitt\,\orcidicon{0000-0001-8173-0862}% <-this % stops a space
    \thanks{B.~H\"afner and G.~A.~Schmitt are with BMW Group Development, 80788 Munich, Germany (e-mail: bernhard.haefner@bmw.de; georg.schmitt@bmw.de).}% <-this % stops a space
    \thanks{J.~Ott is with the Department of Informatics, Technische Universit\"at M\"unchen, 85748 Garching near Munich, Germany (e-mail: ott@in.tum.de).}%
}% <-this % stops a space
\fi

% For IEEE membership, use: \author{IEEE Publication Technology,~\IEEEmembership{Staff,~IEEE,}}

% The paper headers
\ifCLASSOPTIONpeerreview
    \markboth{Vehicular Cooperative Maneuvers - Quo Vaditis}{}
\else
    \markboth{H\"afner \MakeLowercase{\emph{et al.}}: Vehicular Cooperative Maneuvers - Quo Vaditis}
\fi

\maketitle

\begin{abstract}

    Vehicles will not only get more and more automated, but they will also
    cooperate in new ways. Currently, human-driven vehicles begin to
    communicate with each other using \acl*{v2x} technology. Future vehicles
    will use communication to share sensor data and even negotiate cooperative
    maneuvers. This lets them learn more about the environment and improves
    traffic flow and passenger comfort as more predictable maneuvers are likely
    to lead to a smoother ride. This paper introduces the most
    important concepts around cooperative vehicular maneuvers. We also
    summarize currently open challenges and questions to answer before a
    deployment can begin. Afterward, we give some perspectives on the further
    evolution of cooperative maneuvers and beyond. 

\end{abstract}

\begin{IEEEkeywords}
    Cooperative maneuvers, vehicle-to-everything, V2X.
\end{IEEEkeywords}

\section{Introduction}

\IEEEPARstart{I}{n driving} school, besides following driving rules and
controlling a vehicle, instructors teach us to drive cooperatively, for example,
to allow for smooth traffic flow and a pleasant driving experience. For
\acp{adas}, this is similar: while automated and autonomous vehicles can drive
independently, they can leverage their full potential only when cooperating with
each other. 

Cooperation among \acp{cav} has at least three benefits: it mitigates risks
since machines usually function more reliably than humans. It increases driving
comfort by enabling automated vehicles to negotiate maneuvers in advance and
only accepting driving actions that fit the boundary conditions of the driving
state currently chosen by the driver. If drivers want to sleep in their car,
they may be less tolerant of accelerations than if they want to talk via phone.
Lastly, cooperation enables optimized traffic flow since vehicles can find
beneficial maneuvers for themselves and the local vicinity that satisfy all
needs and jointly execute them.

The first step of vehicular communication is broadcasting periodic beacons,
including the vehicle's position and velocity, for basic safety. This direct
\ac{v2x} communication enables several warning use cases independent of
line-of-sight conditions that restrict sensors like cameras or radar. Vehicles
can inform their drivers when a traffic jam starts behind the next road curve.
Or they can warn them of an emergency brake happening several vehicles in front
of them, even if trucks obstruct the direct view. Vehicles can even recognize
when their drivers risk running a red light and warn them accordingly, based on
phase and timing information that intelligent traffic lights can send out. These
scenarios increase drivers' awareness early, potentially saving lives. 

Cooperative maneuvers go beyond such basic safety warning use cases by
transferring decisions to the (automated) vehicle: the \acp{cav} decide on
maneuvers to take, whom to involve, and whether to participate, without human
intervention. 

While prospects are bright, not all issues regarding \ac{ccam} are solved.
Therefore, this paper describes the current status of cooperative maneuver research
and deployments and then sheds light on currently unsolved issues and future
perspectives.  We first introduce the current status of cooperative maneuvers
and what cooperative maneuvers entail and relate to. Then, in
\autoref{sec:challenges}, we show challenges that manufacturers currently face
and which need to be resolved before \acp{oem} can deploy cooperative maneuvers.
Lastly, in \autoref{sec:futureperspectives}, we show future perspectives for
cooperative maneuvers and topics that will gain importance when further
advancing vehicular cooperation. Our earlier survey paper
\cite{Hafner2022-survey} examines the status quo in more depth; we restate and
add the most relevant aspects of cooperative maneuvers in the next section.
Going beyond this survey paper, we concentrate in this publication on challenges
and future perspectives, most of which are different from the open topics
mentioned in the survey paper.

\section{Cooperative Maneuvers - Status Quo and Distinctions}
\label{sec:statusquo}

Cooperative maneuver protocols among \acp{cav} are a relatively new research topic. The
first seminal papers \cite{Hobert2015, Lehmann2018} examining two different
application-layer protocols were published in 2015 and 2018, respectively. 

\begin{figure*}[t]
    \centering
    \subfloat[]{%
        \centering
    \begin{tikzpicture}[line width=1pt, node distance=0.7cm]
        % vehicles
        \node[inner sep=0pt, label={[label distance=-0.20cm]-170:HV}] (hv) at (0,0)
            {\includegraphics[width=1.0cm]{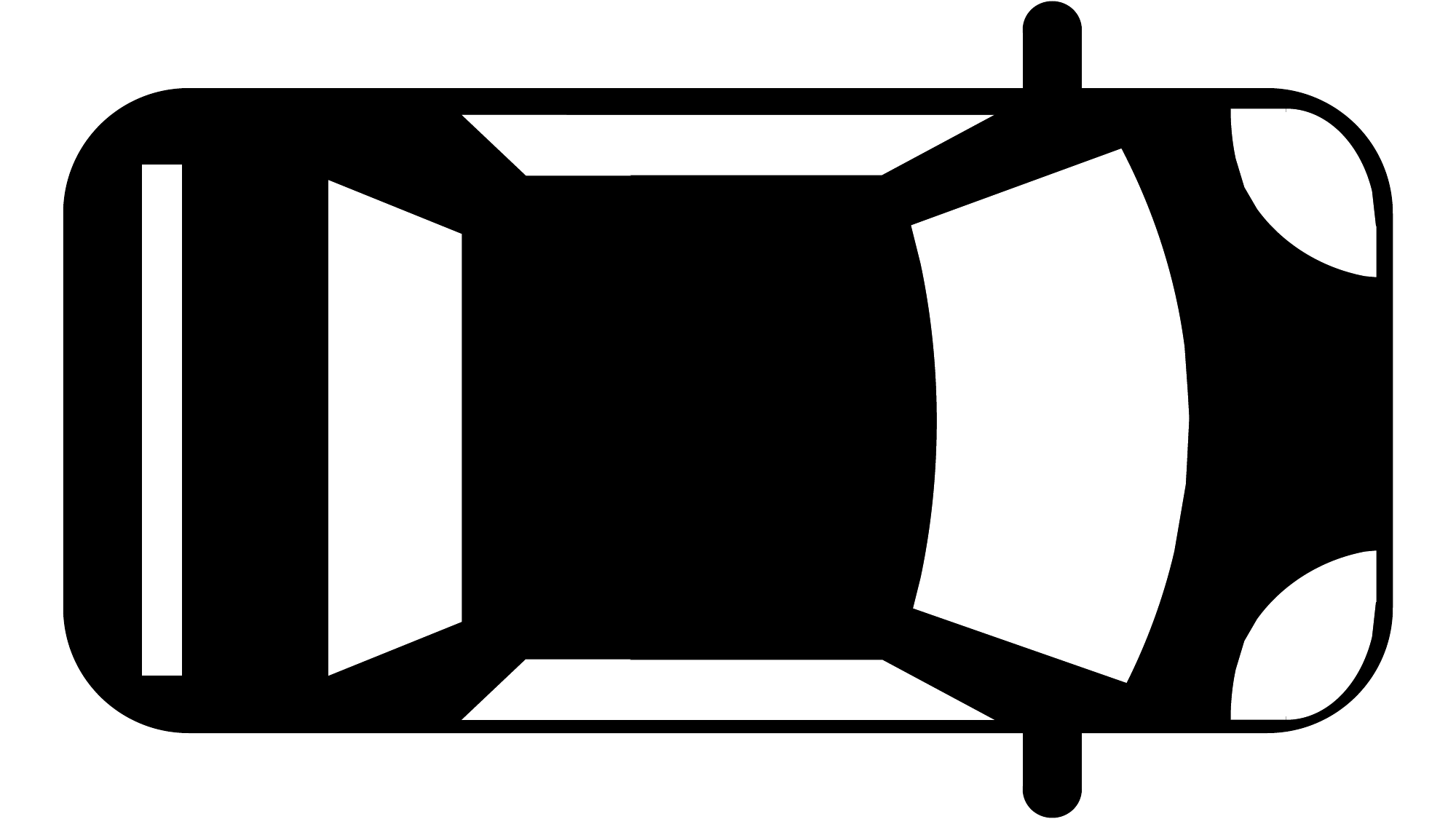}};
        \node[inner sep=0pt, right=0.9cm of hv, label={[label distance=-0.20cm]-170:RV1}] (rv1)
            {\includegraphics[width=1.0cm]{vehicle}};
        \node[inner sep=0pt, above left=0.6cm and 0.1cm of hv, label={[label distance=-0.20cm]-10:RV2}] (rv2)
            {\includegraphics[width=1.0cm]{vehicle}};
        % lanes
        \coordinate [below left=0.2 and 0.6 of hv] (leftend1);
        \coordinate [above left=0.2 and 0.6 of hv] (leftend2);
        \path let \p1=($(leftend2) - (leftend1)$)
            in
            coordinate [above=\y1 of leftend2] (leftend3);
        \draw (leftend1) -- +(0.29\textwidth,0);
        \draw [dashed] (leftend2) -- +(0.29\textwidth,0);
        \draw (leftend3) -- +(0.29\textwidth,0);

        % maneuver arrows
        \draw [mymaneuverarrow] (rv1.east) -- +(0.7,0);
        \draw [mymaneuverarrow] (rv2.east) -- +(3.1,0);
        \draw [mymaneuverarrow] (hv.east) to[out=0,in=-100] ++(0.6,0.5) to[out=80,in=180] ++(0.7,0.5) node (helper) {} ;
        \draw [mymaneuverarrow] (helper) -- ++(0.8,0) node (helper2) {};
        \draw [mymaneuverarrow] (helper2) to[out=0,in=100] ++(0.6,-0.5) to[out=-80,in=180] ++(0.7,-0.5);
    \end{tikzpicture}
    \label{fig:statusquo:usecases:overtake}
    }
    \hfill
    \subfloat[]{%
        \centering
    \begin{tikzpicture}[line width=1pt, node distance=0.7cm]
        % vehicles
        \node[inner sep=0pt, label={[label distance=-0.20cm]-170:HV}] (hv) at (0,0)
            {\includegraphics[width=1.0cm]{vehicle}};
            \node[inner sep=0pt, right=1.3cm of hv, label={[label distance=-0.20cm]-170:RV}] (rv)
            {\includegraphics[width=1.0cm]{vehicle}};
        % lanes
        \coordinate [below left=0.2 and 0.5 of hv] (leftend1);
        \coordinate [above left=0.2 and 0.5 of hv] (leftend2);
        \path let \p1=($(leftend2) - (leftend1)$)
            in
            coordinate [below=\y1 of leftend1] (leftend3);
        \draw (leftend2) -- +(5.5cm,0);
        \draw [dashed] (leftend1) -- +(5.5cm,0);
        \draw (leftend3) -- +(5.5cm,0);

        % maneuver arrows
        \draw [mymaneuverarrow] (hv.east) -- +(1.0,0);
        \draw [mymaneuverarrow] (rv.east) to[out=0,in=100] ++(0.4,-0.5) to[out=-80,in=180] ++(0.5,-0.5) node (helper) {};
        \draw [mymaneuverarrow] (helper) -- +(0.6,0);
    \end{tikzpicture}
    \label{fig:statusquo:usecases:let-pass}
    }
    \hfill
    \subfloat[]{%
        \centering
    \begin{tikzpicture}[line width=1pt, node distance=0.7cm]
        % HV
        \node[inner sep=0pt, label={[label distance=-0.10cm]-180:HV}] (hv) at (0,0)
            {\includegraphics[width=1.0cm]{vehicle}};
        % lanes
        \coordinate [below left=0.05 and 0.5 of hv] (leftend1);
        \coordinate [above left=0.05 and 0.5 of hv] (leftend2);
        \path let \p1=($(leftend2) - (leftend1)$)
            in
            coordinate [above=\y1 of leftend2] (leftend3);
        \coordinate [right=0.29\textwidth of leftend3] (rightend3);
        \draw (leftend1) -- (rightend3 |- leftend1);
        \draw [dashed] (leftend2) -- (rightend3 |- leftend2);
        \draw (leftend3) -- ++(0.1\textwidth,0) node (cornerleft) {} -- ++(0,1cm);
        \draw (rightend3) -- ++(-0.1\textwidth,0) node (cornerright) {} -- ++(0,1cm);
        % RVs
        \node[inner sep=0pt, below left=0.05 and 0.5cm of rightend3, label={[label distance=-0.1cm]0:RV1}] (rv1)
            {\rotatebox[origin=c]{180}{\includegraphics[width=1.0cm]{vehicle}}};
        \node[inner sep=0pt, above right=0.1cm and 0.05cm of cornerleft, label={[label distance=-0.20cm]+10:RV2}] (rv2)
            {\rotatebox[origin=c]{270}{\includegraphics[width=1.0cm]{vehicle}}};
        % maneuver arrows
        \draw [mymaneuverarrow] (rv1.west) -- +(-1.5cm,0);
        \draw [mymaneuverarrow] (rv2.south) arc (180:90:-0.6cm) -- +(-0.5cm,0);
        \draw [mymaneuverarrow] (hv.east) arc (270:360:1.3cm) -- +(0,0.25cm);
    \end{tikzpicture}
    \label{fig:statusquo:usecases:intersection}
    }
    \caption{Three examples for cooperative maneuver use cases. Arrows denote vehicle intents as proposed in a cooperative maneuver. (a) Cooperative overtake, (b) let-pass, (c) intersection crossing}
    \label{fig:statusquo:usecases}
\end{figure*}

The recent SAE standard J3216 proposes classes A through D of cooperative
maneuver protocols, namely \emph{status-sharing}, \emph{intent-sharing},
\emph{agreement-seeking}, and \emph{prescriptive} cooperation.  This distinction
helps discuss and classify cooperative maneuvers in general. Another distinction
within the agreement-seeking (class C) cooperation is between \emph{implicit}
and \emph{explicit} cooperation protocols. With the former, vehicles have to
infer others' plans and subsequent cooperation from responses of indicators like
maneuver options with costs or planned trajectories. If a former conflicting
planned trajectory changes after a vehicle communicates its desire to, e.g.,
change lanes, that vehicle may infer that the vehicle whose trajectory changed
is willing to cooperate. This could fall into Class B cooperation, as long as
only plans are shared. However, many protocols will also share, e.g., desired
alternative trajectories, with the intent that other traffic participants will
accommodate them. This then falls into the category of agreement-seeking
cooperation. In the latter form, vehicles explicitly negotiate on a specific
maneuver. Once they finish negotiating, every involved party can be sure that
everyone agrees on the same maneuver. Examples of what to negotiate include road
space to occupy at a specific time~\cite{Hess2018} or abstract driving action
representations of all involved submaneuvers, e.g., \emph{change speed} or
\emph{change lane}~\cite{Hafner2020-cvip}. The initiating vehicle designs a
maneuver either for itself or including surrounding vehicles. It describes this
maneuver using the building blocks of the respective protocol mentioned above,
e.g., submaneuver descriptions, and transmits it to surrounding vehicles. Once
all to-be participants agree to a proposal, joint maneuver execution can begin.
Several protocols exist in literature~\cite{Hafner2022-survey}, and more
recently research discussion on comparability and relevant metrics
emerged~\cite{Hafner2023-framework}. 

\subsection{Cooperative Maneuvers System Architecture}

There are three important aspects of the architecture for cooperative
maneuvers: \emph{communication} topology, \emph{computation} paradigm, and
the \emph{decision-making} process.
Along all these dimensions, options along a spectrum between
centralized and decentralized are available. All combinations of different
centralization degrees for the overall architecture are possible; however, as of
2023, protocol proposals leverage only a limited set of these options.

\paragraph{Communication topology} Beacon-based approaches like sharing own
trajectories~\cite{Lehmann2018} employ a decentralized communication
topology, meaning all vehicles communicate as equal peers. In contrast, most
explicit maneuver negotiation approaches have a central instance, usually the
initiator of a maneuver with whom all \acp{rv} communicate.

\paragraph{Computation paradigm} Most cooperation protocols distribute the
burden of computation among all (potential) maneuver participants. Every vehicle
has to compute its preferred reaction when receiving an intent for a cooperative
maneuver by another vehicle. Only for a few explicit approaches can the \ac{hv}
compute proposals for the actions of other participants, thus alleviating their
computational burden. This comes at the drawback of having to guess their
preferences.  

\paragraph{Decision-making process} Regarding the decision on what
trajectory to follow, most cooperation protocols assume that the final
responsibility lies with each participating vehicle. Centralized decision-making
could apply to, e.g., \acp{tms} that design optimal behaviors and routes and
would then have to have the authority to prescribe actions to automated
vehicles.

\subsection{Use cases}

The possibilities for using cooperative maneuvers are diverse, cf.
\cref{fig:statusquo:usecases}: vehicles negotiating their driving actions among
each other can align on diverse use cases: cooperatively overtaking a slow
vehicle, making way to let other vehicles pass, or align on how best to cross an
intersection. Also many specialized applications like platooning (decreasing
inter-vehicle distances to a few meters to save energy) are cooperative maneuvers. Researchers and
implementers will potentially discover whole new application areas once
cooperative maneuvers are a reality and all vehicles are automated.

\subsection{The role of infrastructure} \label{sec:statusquo:infrastructure}

Besides traffic participants, traffic infrastructure is the other key element in
a traffic system. Therefore, it is interesting to investigate how to include
roadside facilities in the cooperative maneuver ecosystem. We use the term
\emph{traffic infrastructure} broadly, comprising cameras and other roadside
sensors, \acp{rsu} that communicate information via direct communication, and
mobile communication base stations near roads. 

The first role of such infrastructure is as an information provider:
communication-enabled cameras and \acp{rsu} can inform drivers and automated
vehicles about pedestrians on the street, speed limits, or service areas down
the road. Cameras can also help to include unconnected vehicles by broadcasting
their position and speed information. This makes travel safer by reducing
uncertainty and increasing the available information, e.g., for designing a
cooperative maneuver.

\begin{figure*}
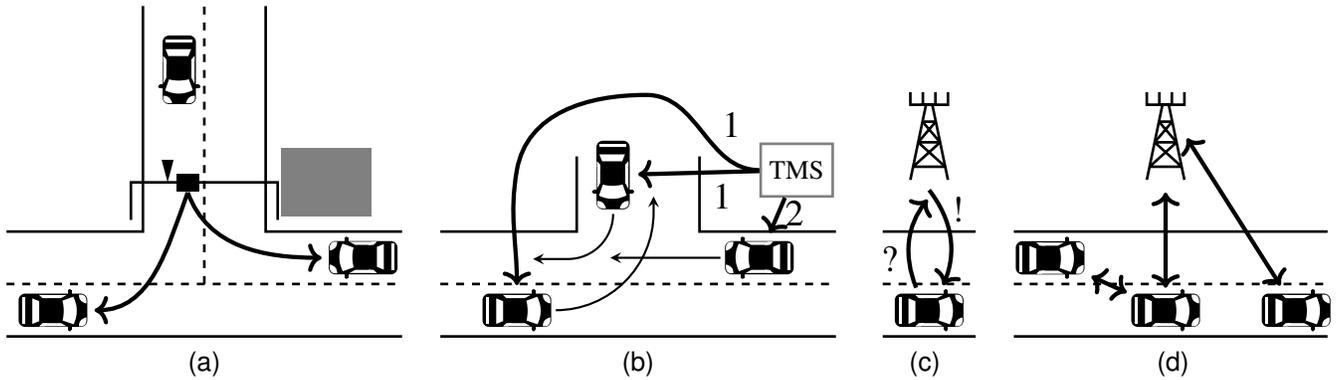

    \subfloat[]{
        \begin{tikzpicture}[line width=1pt, node distance=0.7cm]
            % HV
            \node[inner sep=0pt] (hv) at (0,0)
                {\includegraphics[width=1.0cm]{vehicle}};
            % lanes
            \coordinate [below left=0.05 and 0.1 of hv] (leftend1);
            \coordinate [above left=0.05 and 0.1 of hv] (leftend2);
            \path let \p1=($(leftend2) - (leftend1)$)
                in
                coordinate [above=\y1 of leftend2] (leftend3);
            \coordinate [right=0.29\textwidth of leftend3] (rightend3);
            %\coordinate [right=0.145\textwidth of leftend3] (middle3);
            \draw (leftend1) -- (rightend3 |- leftend1);
            \draw [dashed] (leftend2) -- coordinate [pos=0.5] (middle2) (rightend3 |- leftend2);
            \draw (leftend3) -- ++(0.1\textwidth,0) node (cornerleft) {} --  ++(0,3cm) coordinate (topleft);
            \draw (rightend3) -- ++(-0.1\textwidth,0) node (cornerright) {} -- ++(0,3cm);
            \draw [dashed] (middle2) -- (topleft -| middle2); 
            % RVs
            \node[inner sep=0pt, below left=0.05 and 0.0cm of rightend3] (rv1)
                {\rotatebox[origin=c]{180}{\includegraphics[width=1.0cm]{vehicle}}};
            \node[inner sep=0pt, above right=1.5cm and 0.05cm of cornerleft] (rv2)
                {\rotatebox[origin=c]{270}{\includegraphics[width=1.0cm]{vehicle}}};
            % obstacle
            \node [above right=0.05cm and 0.05cm of cornerright, anchor=south west, fill=gray,minimum width=1.2cm,minimum height=0.9cm] (obstacle) {};
            % CCTV bridge and camera
            \coordinate [above left=0.02cm and 0.025cm of cornerleft] (bridgebottomleft);
            \coordinate [above right=0.02cm and 0.025cm of cornerright] (bridgebottomright);
            \draw (bridgebottomleft) -- ++(0,0.5cm) coordinate (bridgetopleft) -- coordinate[pos=0.25](camerapos) (bridgetopleft -| bridgebottomright) coordinate (bridgetopright) -- (bridgebottomright);
            \fill [fill=black] (camerapos) -- ++(-0.08,0.30) -- ++(0.16,0);
            \node [right=0.1cm of camerapos, anchor=west, fill=black, minimum width=0.3cm, minimum height=0.2cm] (rsu) {}; 
            \draw [mycommunicationarrow] (rsu.south) to[out=290,in=180] (rv1);
            \draw [mycommunicationarrow] (rsu.south) to[out=250,in=0] (hv);
            % maneuver arrows
            % communication arrows
        \end{tikzpicture}
        \label{fig:statusquo:infrastructure:informationprovider}}
    \hfil
    \subfloat[]{
        \begin{tikzpicture}[line width=1pt, node distance=0.7cm]
            % HV
            \node[inner sep=0pt] (hv) at (0,0)
                {\includegraphics[width=1.0cm]{vehicle}};
            % lanes
            \coordinate [below left=0.05 and 0.5 of hv] (leftend1);
            \coordinate [above left=0.05 and 0.5 of hv] (leftend2);
            \path let \p1=($(leftend2) - (leftend1)$)
                in
                coordinate [above=\y1 of leftend2] (leftend3);
            \coordinate [right=0.29\textwidth of leftend3] (rightend3);
            \draw (leftend1) -- (rightend3 |- leftend1);
            \draw [dashed] (leftend2) -- (rightend3 |- leftend2);
            \draw (leftend3) -- ++(0.1\textwidth,0) node (cornerleft) {} -- ++(0,1cm);
            \draw (rightend3) -- ++(-0.1\textwidth,0) node (cornerright) {} -- ++(0,1cm);
            % RVs
            \node[inner sep=0pt, below left=0.05 and 0.5cm of rightend3] (rv1)
                {\rotatebox[origin=c]{180}{\includegraphics[width=1.0cm]{vehicle}}};
            \node[inner sep=0pt, above right=0.1cm and 0.05cm of cornerleft] (rv2)
                {\rotatebox[origin=c]{270}{\includegraphics[width=1.0cm]{vehicle}}};
            % TMS server
            \node [above right=0.5cm and 0.0cm of rv1.north, draw=gray,minimum width=0.7cm,minimum height=0.7cm] (tms) {TMS};
            % maneuver arrows
            \draw [mymaneuverarrow] (rv1.west) -- +(-1.5cm,0);
            \draw [mymaneuverarrow] (rv2.south) arc (180:90:-0.6cm) -- +(-0.5cm,0);
            \draw [mymaneuverarrow] (hv.east) arc (270:360:1.3cm) -- +(0,0.25cm);
            % communication arrows
            \draw [mycommunicationarrow] (tms) -- node [pos=0.4, label={[label distance=-0.2cm]0:{\Large 2}}] {} (rv1);
            \draw [mycommunicationarrow] (tms) -- node [pos=0.3, below] {\Large 1} (rv2);
            \path (hv) |- coordinate(outerright)  (tms);
            \draw [mycommunicationarrow] (tms) to[out=180,in=0] node [pos=0.4, label={[label distance=-0.2cm]80:{\Large 1}}] {} ++(-2cm, 1cm) to[out=180,in=80] (outerright) to[out=260,in=90] (hv);
        \end{tikzpicture}
        \label{fig:statusquo:infrastructure:tms}}
    \hfil
    \subfloat[]{
        \begin{tikzpicture}[line width=1pt, node distance=0.7cm]
            % vehicles
            \node[inner sep=0pt] (hv) at (0,0) 
                {\includegraphics[width=1.0cm]{vehicle}}; 
            % basestation
            \node [basestation, basestation tower height=1.0cm, above=0.4cm of hv] (eNB) {};
            % lanes
            \coordinate [below left=0.05 and 0.1 of hv] (leftend1);
            \coordinate [above left=0.05 and 0.1 of hv] (leftend2);
            \path let \p1=($(leftend2) - (leftend1)$)
                in
                coordinate [above=\y1 of leftend2] (leftend3);
            \coordinate [below right=0.05 and 0.1 of hv] (rightend1);
            \draw (leftend1) -- (rightend1);
            \draw [dashed] (leftend2) -- (rightend1 |- leftend2);
            \draw (leftend3) -- (rightend1 |- leftend3);

            % communication arrows (edge computing)
            \coordinate [above=0.8cm of eNB.south] (endpointrequest);
            \coordinate [above=0.9cm of eNB.south] (startpointresponse);
            %\draw [mycommunicationarrow] (hv) to[bend left] node [pos=0.2, label={[label distance=-0.1cm]300:{\Large ?}}] {} (endpointrequest);
            \draw [mycommunicationarrow] (hv) to[bend left] node [pos=0.3, left] {\Large ?} (endpointrequest);
            \draw [mycommunicationarrow] (startpointresponse) to[bend left] node [pos=0.2, right] {\Large !} (hv);
            \end{tikzpicture}
        \label{fig:statusquo:infrastructure:edgecomputing}}
    \hfil
    \subfloat[]{
        \begin{tikzpicture}[line width=1pt, node distance=0.7cm]
            % vehicles
            \node[inner sep=0pt] (hv) at (0,0) 
                {\includegraphics[width=1.0cm]{vehicle}}; 
            \node[inner sep=0pt, right=0.7cm of hv] (rv1) 
                {\includegraphics[width=1.0cm]{vehicle}}; 
            \node[inner sep=0pt, above left=0.10cm  and 0.5cm of hv] (rv2) 
                {\includegraphics[width=1.0cm]{vehicle}}; 
            % basestation
            \node [basestation, basestation tower height=1.0cm, above=0.4cm of hv] (eNB) {};
            % lanes
            \coordinate [below left=0.05 and 1.5 of hv] (leftend1);
            \coordinate [above left=0.05 and 1.5 of hv] (leftend2);
            \path let \p1=($(leftend2) - (leftend1)$)
                in
                coordinate [above=\y1 of leftend2] (leftend3);
            \coordinate [right=0.23\textwidth of leftend3] (rightend3);
            \draw (leftend1) -- (rightend3 |- leftend1);
            \draw [dashed] (leftend2) -- (rightend3 |- leftend2);
            \draw (leftend3) -- (rightend3);

            % communication arrows (relaying)
            \draw [mycommunicationarrowdouble] (hv) -- (rv2);
            \draw [mycommunicationarrowdouble] (hv) -- ([yshift=0.9cm]eNB.south);
            \draw [mycommunicationarrowdouble] (eNB) -- (rv1);

            \end{tikzpicture}
        \label{fig:statusquo:infrastructure:mediator}}
    \caption{The different roles infrastructure can play within the cooperative maneuver environment. (a) Providing video-captured object information about obstructed vehicles, (b) \acl{tms}, sending an intersection crossing sequence to the vehicles, (c) edge computing resources, receiving a request and sending back the processed results, (d) mediator for vehicles that can only communicate via network-based mobile communications.}
    \label{fig:statusquo:infrastructure}
\end{figure*}

Another role for infrastructure is to function as \acp{tms} \cite{Li2014,
Djahel2015}. These servers optimize traffic on a block-to-city scale unavailable
to single vehicles and then prescribe specific routes or times to cross an
intersection to all relevant vehicles. This increases driving comfort and
reduces emissions from stop-and-go travel, and intersection crossing can itself
be seen as a special cooperative maneuver.

Another role of infrastructure concerns computing. Edge computing
\cite{Mao2017} is a paradigm that offloads computation from end devices, in this
case vehicles, to machines with more computing power that are one or a few
mobile hops away from the device but not as far as a computing cloud. Like this,
\acp{rsu} and mobile base stations can support \acp{cav} by providing resources
for computing driving routes or cooperative maneuvers. It has to be said,
though, that this might pose additional challenges regarding reliability, 
cybersecurity, and deployability of new devices need to be installed. In the
end, if it is guaranteed that \acl{mec} or other suitable
platforms are sufficiently available, manufacturers could reduce the computing
power and thus energy consumption of vehicles, relying on the infrastructure. 

Roadside facilities can finally help to mediate between incompatible actors. It
is unrealistic that competing technologies like 802.11p WiFi-based and
\ac{ltev}-based \ac{v2x} will co-exist in the same region. Nevertheless,
infrastructure can help to mediate between unconnected and connected vehicles or
between different services, as we will further elaborate on in
\cref{sec:challenges:transition}. These different roles show that infrastructure
will forever only be an add-on to onboard computing, since vehicles cannot rely
on wireless communication and available resources.

\subsection{Cooperative maneuver deployments} \label{sec:statusquo:deployments}

Currently, more and more \acp{oem} announce introducing \ac{v2x} services in
their new vehicle models. Especially in China, deployments of \ac{v2x}-based
warnings to drivers take off as \ac{v2x}
is an essential part of the government's \acl{icv} strategy. The deployment of
cooperative maneuvers using \ac{v2x} is still further down the road. Standardization of
cooperative maneuvers has already started: 
the SAE published J3186 describing an explicit negotiation of maneuvers for the United States, and the \ac{etsi} is currently working on
TS~103~561 in Europe, which enables cooperative maneuvers via sharing of planned
and desired trajectories.

\section{Challenges for Cooperative Maneuvers} \label{sec:challenges}

As the previous section shows, research has already addressed many issues
related to cooperative maneuvers. They are on a good path to adoption once
automated vehicles are prevalent. This section sheds light on the issues that
still need to be addressed to bring a working cooperative maneuver system to the
roads.

\subsection{How to transition to cooperative \acp{cav}?}
\label{sec:challenges:transition}

Automation in passenger vehicles is currently very limited. However, most
research on cooperative maneuvers assumes all vehicles are automated and
connected. This gap poses a substantial challenge to the widespread adoption of
cooperative maneuvers. 

\begin{figure*}
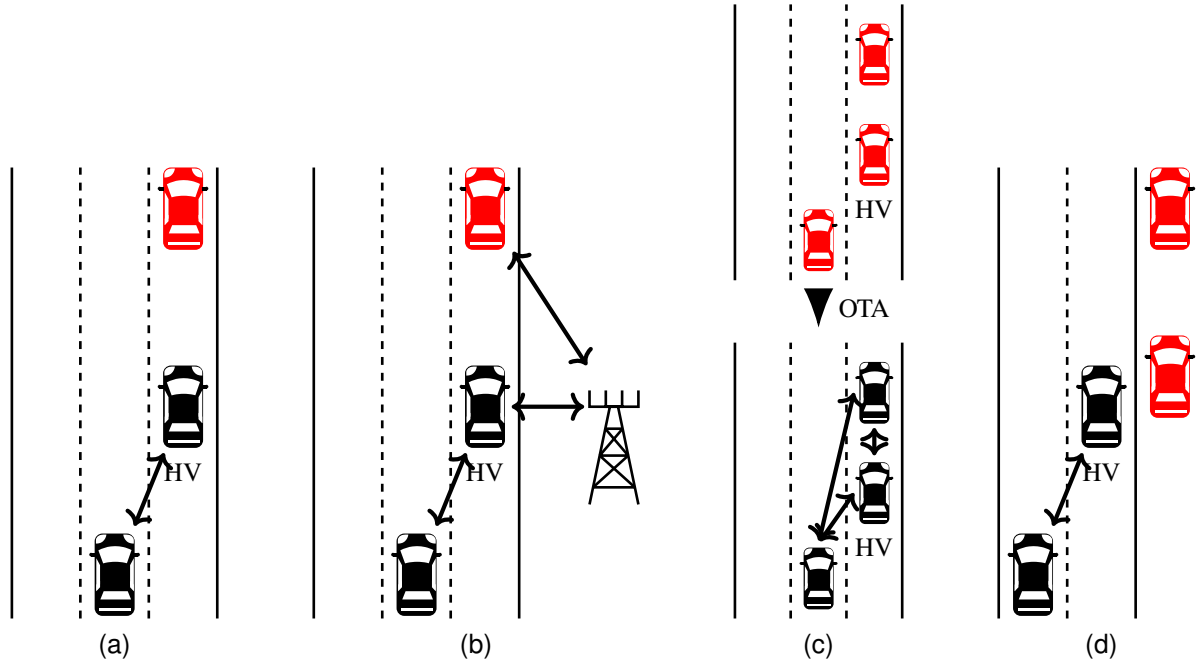

    \centering
    \subfloat[]{
        \begin{tikzpicture}[line width=1pt, node distance=0.7cm]
            % vehicles
            \node[inner sep=0pt, label={[label distance=0.00cm]270:HV}] (hv) at (0,0) 
                {\includegraphics[width=1.2cm, angle=90]{vehicle}}; 
            \node[inner sep=0pt, above=1.4cm of hv] (rv1) 
                {\textcolor{red}{\includegraphics[width=1.2cm, angle=90]{vehicle_color}}}; 
            \node[inner sep=0pt, below left=1.0cm  and 0.2cm of hv] (rv2) 
                {\includegraphics[width=1.2cm, angle=90]{vehicle}}; 
            % lanes
            \coordinate [below left=2.2 and 0.1 of hv] (midrightbottom);
            \coordinate [below right=2.2 and 0.1 of hv] (rightbottom);
            \path let \p1=($(rightbottom) - (midrightbottom)$)
                in
                coordinate [left=\x1 of midrightbottom] (midleftbottom) 
                coordinate [left=\x1 of midleftbottom] (leftbottom);
            \draw (leftbottom) -- +(0, 6cm);
            \draw [dashed] (midrightbottom) -- +(0, 6cm);
            \draw [dashed] (midleftbottom) -- +(0, 6cm);
            \draw (rightbottom) -- +(0, 6cm);
            % communication arrows
            \draw [mycommunicationarrowdouble] (hv) -- (rv2);
            \end{tikzpicture}
        \label{fig:challenges:transition:ignore}}
    \hfil
    \subfloat[]{
        \begin{tikzpicture}[line width=1pt, node distance=0.7cm]
            % vehicles
            \node[inner sep=0pt, label={[label distance=0.00cm]270:HV}] (hv) at (0,0) 
                {\includegraphics[width=1.2cm, angle=90]{vehicle}}; 
            \node[inner sep=0pt, above=1.4cm of hv] (rv1) 
                {\textcolor{red}{\includegraphics[width=1.2cm, angle=90]{vehicle_color}}}; 
            \node[inner sep=0pt, below left=1.0cm  and 0.2cm of hv] (rv2) 
                {\includegraphics[width=1.2cm, angle=90]{vehicle}}; 
            % basestation
            \node [basestation, basestation tower height=1.3cm, right=1.0cm of hv] (eNB) {};
            % lanes
            \coordinate [below left=2.2 and 0.1 of hv] (midrightbottom);
            \coordinate [below right=2.2 and 0.1 of hv] (rightbottom);
            \path let \p1=($(rightbottom) - (midrightbottom)$)
                in
                coordinate [left=\x1 of midrightbottom] (midleftbottom) 
                coordinate [left=\x1 of midleftbottom] (leftbottom);
            \draw (leftbottom) -- +(0, 6cm);
            \draw [dashed] (midrightbottom) -- +(0, 6cm);
            \draw [dashed] (midleftbottom) -- +(0, 6cm);
            \draw (rightbottom) -- +(0, 6cm);
            % communication arrows
            \draw [mycommunicationarrowdouble] (hv) -- (rv2);
            \draw [mycommunicationarrowdouble] (hv) -- (eNB);
            \draw [mycommunicationarrowdouble] (eNB) -- (rv1);
            \end{tikzpicture}
        \label{fig:challenges:transition:hetnet}}
    \hfil
    \subfloat[]{
        \begin{tikzpicture}[line width=1pt, node distance=0.7cm]
            \begin{scope}
                % vehicles
                \node[inner sep=0pt, label={[label distance=0.00cm]270:HV}] (hv) at (0,0) 
                    {\textcolor{red}{\includegraphics[width=0.9cm, angle=90]{vehicle_color}}}; 
                \node[inner sep=0pt, above=0.4cm of hv] (rv1) 
                    {\textcolor{red}{\includegraphics[width=0.9cm, angle=90]{vehicle_color}}}; 
                \node[inner sep=0pt, below left=0.2cm  and 0.2cm of hv] (rv2) 
                    {\textcolor{red}{\includegraphics[width=0.9cm, angle=90]{vehicle_color}}}; 
                % lanes
                \coordinate [below left=1.2 and 0.1 of hv] (midrightbottom);
                \coordinate [below right=1.2 and 0.1 of hv] (rightbottom);
                \coordinate [above right=0.2 and 0.1 of rv1] (righttop);
                \path let \p1=($(rightbottom) - (midrightbottom)$)
                    in
                    coordinate [left=\x1 of midrightbottom] (midleftbottom) 
                    coordinate [left=\x1 of midleftbottom] (leftbottom)
                    coordinate [left=\x1 of righttop] (midrighttop)
                    coordinate [left=\x1 of midrighttop] (midlefttop)
                    coordinate [left=\x1 of midlefttop] (lefttop);
                \draw (leftbottom) -- (lefttop);
                \draw [dashed] (midrightbottom) -- (midrighttop);
                \draw [dashed] (midleftbottom) -- (midlefttop);
                \draw (rightbottom) -- (righttop);
                % communication arrows
            \end{scope}
            \coordinate [below=0.2 of rv2] (transition);
            \draw [double distance=3pt, arrows = {-Latex[length=0pt 3 0]}] (transition) -- +(0,-0.5) node [label={[label distance=0.0]5:OTA}] {};
            \begin{scope}[yshift=-4.5cm]
                % vehicles
                \node[inner sep=0pt, label={[label distance=0.00cm]270:HV}] (hv) at (0,0) 
                    {\includegraphics[width=0.9cm, angle=90]{vehicle}}; 
                \node[inner sep=0pt, above=0.4cm of hv] (rv1) 
                    {\includegraphics[width=0.9cm, angle=90]{vehicle}}; 
                \node[inner sep=0pt, below left=0.2cm  and 0.2cm of hv] (rv2) 
                    {\includegraphics[width=0.9cm, angle=90]{vehicle}}; 
                % lanes
                \coordinate [below left=1.2 and 0.1 of hv] (midrightbottom);
                \coordinate [below right=1.2 and 0.1 of hv] (rightbottom);
                \coordinate [above right=0.2 and 0.1 of rv1] (righttop);
                \path let \p1=($(rightbottom) - (midrightbottom)$)
                    in
                    coordinate [left=\x1 of midrightbottom] (midleftbottom) 
                    coordinate [left=\x1 of midleftbottom] (leftbottom)
                    coordinate [left=\x1 of righttop] (midrighttop)
                    coordinate [left=\x1 of midrighttop] (midlefttop)
                    coordinate [left=\x1 of midlefttop] (lefttop);
                \draw (leftbottom) -- (lefttop);
                \draw [dashed] (midrightbottom) -- (midrighttop);
                \draw [dashed] (midleftbottom) -- (midlefttop);
                \draw (rightbottom) -- (righttop);
                % communication arrows
                \draw [mycommunicationarrowdouble] (hv.west) -- (rv2.north);
                \draw [mycommunicationarrowdouble] (rv1.west) -- (rv2.north);
                \draw [mycommunicationarrowdouble] (rv1) -- (hv);
            \end{scope}
        \end{tikzpicture}
        \label{fig:challenges:transition:ota}}
    \hfil
    \subfloat[]{
        \begin{tikzpicture}[line width=1pt, node distance=0.7cm]
            % vehicles
            \node[inner sep=0pt, label={[label distance=0.00cm]270:HV}] (hv) at (0,0) 
                {\includegraphics[width=1.2cm, angle=90]{vehicle}}; 
            \node[inner sep=0pt, above right=1.4cm and 0.2cm of hv] (rv1) 
                {\textcolor{red}{\includegraphics[width=1.2cm, angle=90]{vehicle_color}}}; 
            \node[inner sep=0pt, below left=1.0cm  and 0.2cm of hv] (rv2) 
                {\includegraphics[width=1.2cm, angle=90]{vehicle}}; 
            \node[inner sep=0pt, below=1.0cm of rv1] (rv3) 
                {\textcolor{red}{\includegraphics[width=1.2cm, angle=90]{vehicle_color}}}; 
            % lanes
            \coordinate [below left=2.2 and 0.1 of hv] (midleftbottom);
            \coordinate [below right=2.2 and 0.1 of hv] (midrightbottom);
            \path let \p1=($(midrightbottom) - (midleftbottom)$)
                in
                coordinate [left=\x1 of midleftbottom] (leftbottom) 
                coordinate [right=\x1 of midrightbottom] (rightbottom);
            \draw (leftbottom) -- +(0, 6cm);
            \draw (midrightbottom) -- +(0, 6cm);
            \draw [dashed] (midleftbottom) -- +(0, 6cm);
            \draw (rightbottom) -- +(0, 6cm);
            % communication arrows
            \draw [mycommunicationarrowdouble] (hv) -- (rv2);
            \end{tikzpicture}
        \label{fig:challenges:transition:separation}}
        \caption{Different transition scenarios towards every vehicle participating in cooperative maneuvers; black vehicles can perform cooperative maneuvers via direct communication, red vehicles cannot . (a) Treating incompatible vehicles as obstacles, (b) inclusion via edge relays, (c) collective over-the-air update, (d) separate lanes for \acp{cav}.  }
    \label{fig:challenges:transition}
\end{figure*}

Some funded projects have looked into the scenario of mixed traffic, like the
InfraMix and
TransAID\footnote{See \href{www.transaid.eu}{www.transaid.eu}} projects in
Europe. The former designed infrastructure that supports automated and
non-automated vehicles for scenarios like bottlenecks, road works, or dynamic
lane assignments, including speed recommendations. Likewise, TransAID designed a
solution for infrastructure-led assignment of automation levels to facilitate
the co-existence of automated, connected, and neither automated nor connected
vehicles. Such projects have started to investigate the issue of mixed traffic.
In the following and in \autoref{fig:challenges:transition}, we show more
specific ideas of how a transition from vehicles unable to participate in
cooperative maneuvers via vehicular communication to \acp{cav} that are could
happen.

\paragraph{Enable only new vehicles} If relying solely on direct communication, the first vehicles introducing the
capability for cooperative maneuvers will not be able to use them often -- they
will not meet other enabled vehicles. In contrast to basic \ac{v2x}
communication, where infrastructure-based use cases can provide value-add from
the first vehicle onwards, cooperative maneuvers always require all involved
vehicles to have the respective capabilities. However, regarding new traffic
infrastructure including \acp{rsu}, municipalities usually have to
consider the value-add they deploy, which is small when only a few
vehicles can use the information. So there exists a chicken-and-egg problem for
infrastructure, as well. 

\paragraph{Relay messages} A second option could be that \ac{oem} backends or edge servers facilitate
cooperative maneuvers by relaying messages,  as a mobile (backend) connection is
already widely available in today's vehicles. \acp{oem} usually prefer running
services via their own backends. However, the communication chain via several
backends may not be feasible to coordinate cooperative maneuvers in the expected
time frame. Edge servers could help coordinate cooperative maneuvers locally by
relaying messages to nearby vehicles, cf.
\cref{fig:challenges:transition:hetnet}. An option worth considering in case
cooperative maneuvers via mobile network connectivity are possible is whether
this network-based, as opposed to direct, communication should not be the
standard operating mode for \acp{cav} performing cooperative maneuvers. This
would eliminate the need for extra communication hardware, and, at least in
theory, all vehicles with backend connectivity would have the opportunity to
participate in cooperative maneuvers. However, the savings in onboard hardware must be
weighed against the increased mobile connectivity costs. Latency-wise, we have
proposed in an earlier publication\cite{Hafner2022-failures} that the critical factor for
cooperative maneuvers is a negotiation duration of below \SI{1}{\second}, since
during this time frame, traffic situations may not change too much.

\paragraph{\Acl{ota} updates} Another way would be to activate the capabilities after purchase via \ac{ota}
updates once a critical mass of `latently-enabled' vehicles is driving on the
roads, cf. \cref{fig:challenges:transition:ota}. However, this possibility needs
to be considered by vehicle \acp{oem} very early on, as it requires projecting
vehicle capabilities several years into the future at a time when not even
automation is that widespread. They then must ensure sufficient computing
resources, suitable in-car networks, and other details parallel to increasing
demands from entertainment and infotainment services.

\paragraph{Dedicated lanes for \acp{cav}} The above three options are all related to the technology present in the
vehicles. An option open to road authorities in cooperation with vehicle
manufacturers is to dedicate specific lanes to \acp{cav} only, cf.
\cref{fig:challenges:transition:separation}. Specific highway lanes are already
reserved for automated vehicles that do not require human interaction, e.g., on
the Jingxiong highway in China.\footnote{Cf.
\href{https://bit.ly/ADHWeng}{https://bit.ly/ADHWeng} (in English, from May
2021) or \href{https://bit.ly/ADHWChin}{https://bit.ly/ADHWChin} (in Chinese,
from April 2023).} Something similar could be done with (at least two adjacent)
lanes that are open only to vehicles capable of cooperative driving. The first
vehicles would then enjoy practically free travel, further incentivizing
producing and purchasing \acp{cav}. However, the feasibility of this option
depends on variables such as future traffic volume in a region, synchronization
of \ac{oem} starts of production of the cooperative driving feature, and law
enforcement powers to control adherence to such restrictions. This seems 
hard as it would probably cause traffic jams right at the outset when road
infrastructure is scarce and this solution effectively reduces the amount of
available lanes for legacy vehicles, which will make up the majority on the road
for some time.

\subsection{How to ensure sufficient security?}

For basic safety, a \ac{pki} system that uses certificates to sign \ac{v2x}
messages already realizes \ac{v2x} security and trust \cite{Brecht2018}. While
regional variations exist, stakeholders globally agree on this \ac{pki} system
to provide sufficient security and trust for applying \ac{v2x}. Vehicles can
verify the authenticity of the received messages and be sure they stem from
legitimate traffic entities like traffic lights or other vehicles.

However, a big premise for \ac{v2x} basic safety use cases is that they are only
used to warn a human driver. No automated driving maneuver is based solely on
received \acp{bsm} or triggered warnings. At least, vehicles will cross-check
\ac{v2x} warnings with inputs from other sensors to validate, e.g., whether
there is a traffic jam ahead. For cooperative maneuvers, this drastically
changes as the default mode is to negotiate and execute maneuvers among
\acp{cav}. Even more so, since vehicles negotiate their \emph{future} driving
actions, validating the plausibility of the \ac{v2x} message contents using
other sensors is only possible up to the extent of using current sensor inputs
and projecting into the future.  

Therefore, it is likely that going beyond the \ac{v2x} \ac{pki} system,
additional security mechanisms need to be put in place to ensure that
cooperative maneuvers are safe to conduct. Only if standards achieve sufficient
\ac{sotif} then \acp{oem} and customers will want to implement and use
cooperative maneuvers in their vehicles.

Up to now and to the best of the authors' knowledge, there are no peer-reviewed
studies on security for cooperative maneuvers. Hence, the first fundamental open
issue is to define what ``sufficient'' security for cooperative maneuvers even
means. With clearly defined security goals, experts can analyze the \ac{v2x}
\ac{pki} and identify needs and solutions for additional security measures.

\subsection{Legal considerations}

Up to now, mainly technology drives progress for \ac{v2x} applications, while
the legal aspects of \ac{v2x} have been mostly neglected.

For basic safety, drivers could claim being distracted by \ac{v2x} warnings and
thus losing control over a usually controllable situation. Overall, the human
driver stays in charge. 

With cooperative maneuvers, the same liability issues arise as with autarkic
\acp{adas}, namely how far the liability of the \ac{oem} or even the software
development teams go instead of the person being driven in the car. Going beyond
single-vehicle \ac{adas}, lawyers should also investigate and discuss the
liability in case that vehicle A performs actual driving actions solely based on
message exchanges with vehicle B.

We can draw an analogy to level~4 automated driving during the \ac{avp} use
case, where the legal situation (in Germany as an example) is clarified since a
publication from October~2022: the Federal Motor Transport Authority
(\emph{Kraftfahrt-Bundesamt}, KBA) released a requirements catalog for
\ac{avp}~\cite{KBAAVP2022}. Suppose vehicle \acp{oem} fulfill the requirements
in this catalog, e.g., regarding validation of simulation, real-world test
cases, or the design and usage of a security concept. In that case, they will be
allowed to deploy this \ac{avp} function. In type~2 \ac{avp}, the vehicle will
just perform driving actions provided by the infrastructure. If an
accident still happens while the \ac{avp} system is operating, the
infrastructure provider would be liable, not the vehicle
\ac{oem}. However, how such liability issues will apply to \acp{cav} with
level~3 or higher, where the driver is not responsible for driving any more, is
currently unclear. The legal details of this setup  need to be discussed
further.

Something similar is also necessary for cooperative maneuvers. Experts are
debating which protocols may lend themselves most readily for functional safety
analysis, implicit or explicit ones. Some standardization bodies also started
working on functional safety for automated driving and cooperative maneuvers. In
essence, cooperative maneuvers try to establish a consensus in a network of
untrustworthy, distributed actors. This problem of \emph{Byzantine fault
tolerance} is studied extensively among computer networks\cite{Distler2021}, so
potentially findings from this field may help to analyze functional safety for
cooperative maneuver protocols. Likewise, the available research corpus on
Byzantine fault tolerance may help refine the cooperative maneuver protocols to
account for arbitrary failures of the participating vehicles. Once a rigorous
functional safety analysis for cooperative maneuvers is finished, traffic
authorities may use this analysis to create a legal framework around cooperative
maneuvers. Automotive \acp{oem} could then also get certified on their
cooperative maneuver systems, attesting their functional safety standard.

Designing such a legal framework will draw upon technical evaluations of testing
versus verification. It still needs to be determined how much of both is
necessary and sufficient or whether it is even possible to sufficiently cover
all scenarios before a cooperative system would be allowed onto the road. 

\subsection{How to adapt \acs{v2x} technology to cooperative maneuvers?}

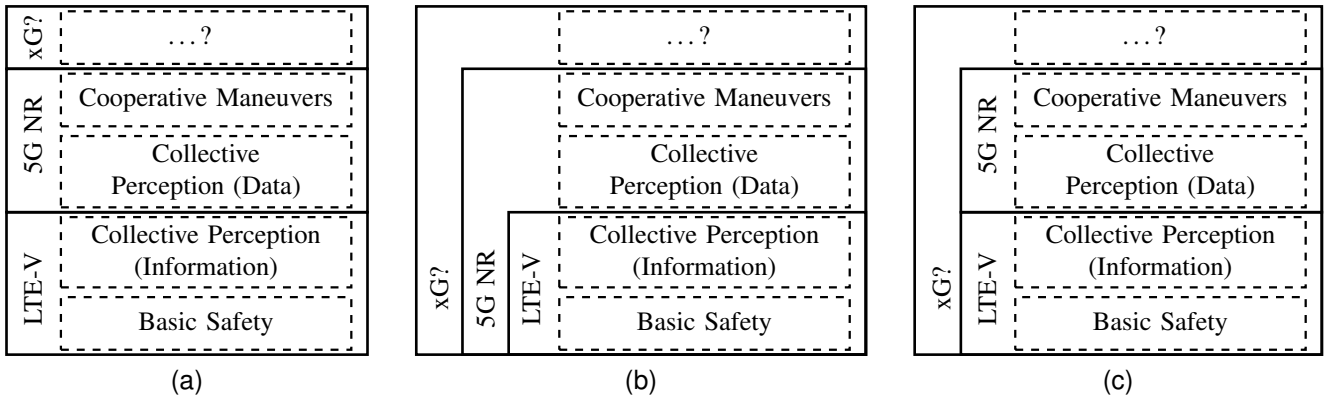
\begin{figure*}
    \subfloat[]{
        \begin{tikzpicture}[line width=1pt, node distance=0.7cm]
            % Services
            \node [mybox, minimum height=0.7cm, minimum width=3.7cm, text width=3.65cm, dashed] (basicsafety) at (0,0) {Basic Safety};
            \node [mybox, minimum height=0.7cm, minimum width=3.7cm, text width=3.65cm, dashed, above=0.1cm of basicsafety] (collpercinfo) {Collective Perception (Information)};
            \node [mybox, minimum height=0.7cm, minimum width=3.7cm, text width=3.65cm, dashed, above=0.1cm of collpercinfo] (collpercdata) {Collective Perception (Data)};
            \node [mybox, minimum height=0.7cm, minimum width=3.7cm, text width=3.65cm, dashed, above=0.1cm of collpercdata] (coopman) {Cooperative Maneuvers};
            \node [mybox, minimum height=0.7cm, minimum width=3.7cm, text width=3.65cm, dashed, above=0.1cm of coopman] (next) {\dots?\phantom{Ag}};
            % Technologies
            \node [above left=0.1cm and 0.1cm of basicsafety, rotate=90, anchor=south] (ltev) {\acs{ltev}};
            \node [below left=0.1cm and 0.1cm of coopman, rotate=90, anchor=south] (5gnr) {\acs{5g} NR};
            \node [left=0.1cm of next, rotate=90, anchor=south] (xg) {xG?};
            % Helper coordinates
            \path (ltev.north west) |- coordinate (bottomleft) (basicsafety.south)  ;
            \path (5gnr.north east) |- coordinate (helper) (coopman.north)  ;
            \path (helper) -| coordinate (topleft) (bottomleft)  ;
            % Boxes
            \draw [draw=black] ([xshift=-0.1cm,yshift=-0.05cm]bottomleft) rectangle ([xshift=0.2cm,yshift=0.05cm]collpercinfo.north east);
            \draw [draw=black] ([xshift=-0.1cm,yshift=0.05cm]topleft) rectangle ([xshift=0.2cm,yshift=-0.05cm]collpercdata.south east);
            \draw [draw=black] ([xshift=-0.1cm,yshift=0.05cm]topleft) rectangle ([xshift=0.2cm,yshift=0.05cm]next.north east);
        \end{tikzpicture}
        \label{fig:perspectives:adaptation:newtech}}
    \hfil
    \subfloat[]{
        \begin{tikzpicture}[line width=1pt, node distance=0.7cm]
            % Services
            \node [mybox, minimum height=0.7cm, minimum width=3.7cm, text width=3.65cm, dashed] (basicsafety) at (0,0) {Basic Safety};
            \node [mybox, minimum height=0.7cm, minimum width=3.7cm, text width=3.65cm, dashed, above=0.1cm of basicsafety] (collpercinfo) {Collective Perception (Information)};
            \node [mybox, minimum height=0.7cm, minimum width=3.7cm, text width=3.65cm, dashed, above=0.1cm of collpercinfo] (collpercdata) {Collective Perception (Data)};
            \node [mybox, minimum height=0.7cm, minimum width=3.7cm, text width=3.65cm, dashed, above=0.1cm of collpercdata] (coopman) {Cooperative Maneuvers};
            \node [mybox, minimum height=0.7cm, minimum width=3.7cm, text width=3.65cm, dashed, above=0.1cm of coopman] (next) {\dots?\phantom{Ag}};
            % Technologies
            \node [above left=0.1cm and 0.1cm of basicsafety, rotate=90, anchor=south] (ltev) {\acs{ltev}};
            \node [left=0.1cm of ltev.north, rotate=90, anchor=south] (5gnr) {\acs{5g} NR};
            \node [left=0.1cm of 5gnr.north, rotate=90, anchor=south] (xg) {xG?};
            % Helper coordinates
            \path (ltev.north west) |- coordinate (bottomleft) (basicsafety.south)  ;
            \path (5gnr.north west) |- coordinate (helper) (coopman.north)  ;
            \path  (helper) |- coordinate (outerleft) (bottomleft);
            \path  (outerleft) -| coordinate (helper2) (xg.north west);
            \path  (helper2) |- coordinate (faroutleft) (bottomleft);
            % Boxes
            \draw [draw=black] ([xshift=-0.05cm,yshift=-0.05cm]bottomleft) rectangle ([xshift=0.2cm,yshift=0.05cm]collpercinfo.north east);
            \draw [draw=black] ([xshift=-0.05cm,yshift=-0.05cm]outerleft) rectangle ([xshift=0.2cm,yshift=0.05cm]coopman.north east);
            \draw [draw=black] ([xshift=-0.05cm,yshift=-0.05cm]faroutleft) rectangle ([xshift=0.2cm,yshift=0.05cm]next.north east);
        \end{tikzpicture}
        \label{fig:perspectives:adaptation:backward}}
    \hfil
    \subfloat[]{
        \begin{tikzpicture}[line width=1pt, node distance=0.7cm]
            % Services
            \node [mybox, minimum height=0.7cm, minimum width=3.7cm, text width=3.65cm, dashed] (basicsafety) at (0,0) {Basic Safety};
            \node [mybox, minimum height=0.7cm, minimum width=3.7cm, text width=3.65cm, dashed, above=0.1cm of basicsafety] (collpercinfo) {Collective Perception (Information)};
            \node [mybox, minimum height=0.7cm, minimum width=3.7cm, text width=3.65cm, dashed, above=0.1cm of collpercinfo] (collpercdata) {Collective Perception (Data)};
            \node [mybox, minimum height=0.7cm, minimum width=3.7cm, text width=3.65cm, dashed, above=0.1cm of collpercdata] (coopman) {Cooperative Maneuvers};
            \node [mybox, minimum height=0.7cm, minimum width=3.7cm, text width=3.65cm, dashed, above=0.1cm of coopman] (next) {\dots?\phantom{Ag}};
            % Technologies
            \node [above left=0.1cm and 0.1cm of basicsafety, rotate=90, anchor=south] (ltev) {\acs{ltev}};
            \node [below left=0.1cm and 0.1cm of coopman, rotate=90, anchor=south] (5gnr) {\acs{5g} NR};
            \node [left=0.1cm of ltev.north, rotate=90, anchor=south] (xg) {xG?};
            % Helper coordinates
            \path (ltev.north west) |- coordinate (bottomleft) (basicsafety.south)  ;
            \path (5gnr.north east) |- coordinate (helper) (coopman.north)  ;
            \path (helper) -| coordinate (topleft) (bottomleft)  ;
            \path  (bottomleft) -| coordinate (helper2) (xg.north west);
            \path  (helper2) |- coordinate (faroutleft) (bottomleft);
            % Boxes
            \draw [draw=black] ([xshift=-0.1cm,yshift=-0.05cm]bottomleft) rectangle ([xshift=0.2cm,yshift=0.05cm]collpercinfo.north east);
            \draw [draw=black] ([xshift=-0.1cm,yshift=0.05cm]topleft) rectangle ([xshift=0.2cm,yshift=-0.05cm]collpercdata.south east);
            \draw [draw=black] ([xshift=-0.1cm,yshift=-0.05cm]faroutleft) rectangle ([xshift=0.2cm,yshift=0.05cm]next.north east);
        \end{tikzpicture}
        \label{fig:perspectives:adaptation:hybrid}}
        \caption{Different scenarios for technology adaptation for cooperative maneuvers. (a) New technologies for new services, (b) new technologies also supporting old services, (c) hybrid model.}
    \label{fig:perspectives:adaptation}
\end{figure*}

It is clear that cooperative maneuvers are the next, but certainly not the last
evolutionary step for vehicular cooperation. Therefore, it is important to
decide on a method for extending \ac{v2x} for future services, especially as the
transmission technology should generally fit the semantics of the intended
application. Even if cooperative maneuvers themselves do not require more
advanced technology than \ac{ltev} or IEEE 802.11p, the question of technology
evolution will arise, and implementers should discuss the appropriate approach.
Ideally, all regions deploying \ac{v2x} globally would align on a common
technological evolution path.  
\cref{fig:perspectives:adaptation} visualizes the three options described in the
following.

The first method states, ``new services with new technologies,'' tailoring
additional technologies for the needs of new future services. This would entail
that new vehicles always incorporate all prior \ac{v2x} technologies, increasing
the overall range of used technologies. For example, 5G New Radio \ac{v2x} is
incompatible on the physical layer with \ac{ltev}. This would mean either
additional hardware for each technology or---where possible---reusing existing
hardware like antennas and somehow multiplexing the resources in, e.g., time.
Both aspects may be possible with two technologies like 802.11p or \ac{ltev} for
basic safety and 5G sidelink for cooperative maneuvers. However, they become
increasingly challenging with additional third or fourth technologies to support
future services. This method could also create lock-ins of technology far into
the future, and design errors in old technology would either need to be accepted
or fixed years or decades after the technology should have been finalized.

The second approach involves new technologies for old and new services and
making them backward compatible, as seen with the 802.11bd efforts. New vehicles
would then need only to incorporate one technology. Only the issue of
multiplexing in time, frequency, or other resources remains. Regarding lock-in,
if newer vehicles only support new technologies, they may slowly replace the
older technologies. However, new technologies can then never truly be
independent of design errors in past technologies, as they need to be compatible
with them. The old technologies will phase out over time, but this will take a
long time, especially given vehicles' long average life spans. Therefore, future
cars and roadside infrastructure would need to be backwards compatible for
sufficiently long.

Combining the two approaches may be a compromise: design new technologies for
old and new services that are backward compatible, up to some point where the
communications and automotive industries jointly decide to break with the past
and design the next technology so that it better enables all old services at the
cost of breaking compatibility with the prior technology. For a specific period,
vehicles would need to incorporate both the prior and the subsequent technology
until a sufficiently high percentage of vehicles supports the new technology and
the prior one can be discontinued. One real-world example is the persistence of
the 2G mobile standard, even in times of 5G.

One such incompatible technology shift could be that some future backend
architecture and communication technology, like a hypothetical 8G, may enable
all \ac{v2x} services. New vehicles supporting 8G plus direct communication
could then serve to phase out direct communication. After the transition period
of several years, new vehicles would only incorporate 8G backend connectivity
and still support all \ac{v2x} services.

Another issue related to the technology choice for \ac{v2x} applications is
spectrum availability. Vehicles are only allowed to send \ac{v2x} messages in a
very restricted communications band, and most available bands are already
reserved for certain types of communication. Additionally, if spectrum frees up,
various parties often dispute why their technology or use case should be given
new spectrum resources. Whether future \ac{v2x} services that need additional
spectrum will get enough bandwidth is currently unclear.

\section{Future Perspectives} \label{sec:futureperspectives}

In contrast to the last section that summarized challenges for which the
community needs to find solutions \emph{before} cooperative maneuvers can be
deployed, this section mentions some aspects that will become relevant
\emph{after} cooperative maneuvers will have appeared in customer vehicles on
the road.

\subsection{How can different cooperative maneuver protocols co-exist?}

Regions worldwide harmonized their approach to basic safety. Vehicles
periodically send beacons, including basic information like position, velocity,
and heading, and receiving vehicles can trigger warnings based on them. The
concrete implementation varies slightly, e.g., between the SAE \acp{bsm} in the
US and the European \ac{cam}. Likewise, regional variation in cooperative
maneuver protocols is likely, probably with bigger differences than between
\acp{bsm} and \acp{cam}. Standardization efforts for cooperative maneuver
protocols in Europe and the United States already differ. While the former will
likely use an implicit approach sharing trajectories, the latter standard is
based on explicitly negotiating for road space. 

Besides regional variations, there will likely not be a `one-size-fits-all'
protocol for all applications. Research into which metrics are suitable for
comparing cooperation protocols has just started, and highly-specialized
applications like platooning or intersection management will likely benefit from
specialized protocols. On the other hand, a generally applicable cooperative
maneuver protocol could cover a wide range of standard maneuvers. These groups
are, however, not entirely distinct, and there will be overlaps where different
protocols can enable the same use case. It needs to be determined whether this
is an advantage since it creates fallback possibilities for maneuver negotiation
or whether it creates confusion and is thus an issue to avoid. In any case, it
needs careful design of standards and onboard algorithms.

For the time being, it could be a good idea to use intent-sharing as a common
platform that every vehicle understands. Even if some vehicles cannot
participate in more complex maneuver negotiations, because they do not support a
certain version of a protocol, they can still share their intents including the
next planned driving actions within a time horizon of, e.g., \num{3} to
\SI{5}{\second}. With intent-sharing as smallest interoperable unit, all
vehicles could enjoy an enriched pool of information about the environment,
potentially ensuring big parts of the additional gains in traffic flow and
driver comfort.

\subsection{How to make intelligent decisions on cooperative maneuvers} 

A related issue concerns how to decide intelligently on cooperative maneuvers.
With the advancement of \acl{ai}, possibilities to determine the next maneuvers,
purely own or in cooperation with others, will increase. Future algorithms will
not only have to decide which trajectory to follow and what driving action to
perform, but they have the additional possibility to involve others to increase
overall utility across all involved parties. They have to weigh this benefit
against the increased negotiation efforts, probably depending also on traffic
and communication channel conditions and the available cooperation protocol(s).

\subsection{How to go beyond cooperative maneuvers?} 

The political vision for \ac{ccam} seems clear: to combine vehicles' communication,
automation, and cooperation capabilities to create an environment where all
traffic participants cooperate on all matters all the time. Not in distinct
instances like a cooperative maneuver or an instance of collective perception,
but rather continuously throughout the time when vehicles drive within each
other's vicinity. 

However, from a technological perspective, this roadmap could be clearer.
Increasing levels of automation and improved communication technologies are
relatively straightforward. However, some open questions exist regarding
cooperation: cooperation progressed from sharing basic status information in
\acp{bsm} or \acp{cam}, to sharing sensor information, to aligning cooperative
maneuvers. However, the next step must be made clear: how to progress from
distinct maneuvers to continuous cooperation? Potentially completely new
protocols would be necessary for this. 

One evolutionary way to transition is by improving the application algorithms
responsible for identifying cooperation possibilities to find more situations
worthy of cooperation. With general-purpose cooperation protocols, this would be
possible in theory. Researchers may also propose extensions to existing
cooperative maneuver protocols to allow, e.g., concurrent maneuvers or updating
a maneuver while it is in progress. 

Another more revolutionary option is to design entirely new protocols for
continuous cooperation -- potentially incorporating the sensor sharing and
cooperative maneuver applications more efficiently. While this approach could
enable continuous cooperation in all situations, a study has yet to propose such
a new protocol, indicating that its introduction may still be decades away. A
new protocol, especially if it incorporates the older applications of sensor
sharing and cooperative maneuvers, will also face the same transition and
ramp-up issues we identified in \cref{sec:challenges:transition}.

When expanding the range of deployed use cases, one aspect to consider is the
legacy on the road. When introducing cooperative maneuvers, vehicles sold today
will become legacy, as they do not yet support them, cf.
\cref{fig:challenges:transition}. When adding additional use cases within or
beyond single cooperative maneuvers, tomorrow's vehicles will also quickly
become outdated. Therefore, deployments of cooperative maneuvers should consider
how to make the system adaptive to either update vehicles in the field,
potentially beyond current over-the-air upgrade capabilities, or provide
backward compatibility mechanisms for future vehicles. With an increasing range
of interactions via \ac{v2x}, feature clashes could also become an increasingly
important issue: implementers need to ensure that new features of cooperative
maneuvers or beyond are compatible with all existing features. After all,
cooperative maneuvers and continuous cooperation are an optimization, and
vehicles will still need to function even without cooperation. Potentially, the
feasible and most suitable operational point lies between the two extremes of
never and always cooperating.

While single cooperative maneuvers are, per definition, only concerned about the
immediate vicinity of the participating vehicles, a continuous cooperation
approach may have more far-reaching consequences. Notably, the range of
information from continuous cooperation may increase, e.g., to propagate
information on traffic conditions further down the road. Such a spreading of
information may or may not be beneficial compared to either not having such
multi-hop information spreads or to providing traffic information via, e.g.,
sensors, \acp{rsu}, and other road infrastructure. 

\section{Conclusion}

This paper gave an overview of cooperative maneuvers for \acp{cav}. The readers
now know about protocols, architecture aspects, and infrastructure involvement options
for cooperative maneuvers. They are aware of current challenges and relevant
topics once cooperative maneuvers are deployed on the roads. We hope this paper
contributes to more researchers picking up interest in the topic, thus
facilitating further analysis and advances in the future.

%%%%%%%%%%%%%%%%% End of content 
% References
\bibliography{outlook}
\bibliographystyle{IEEEtran}

% biography section
\begin{IEEEbiography}
    [{\includegraphics[width=1in,height=1.25in,clip,keepaspectratio]{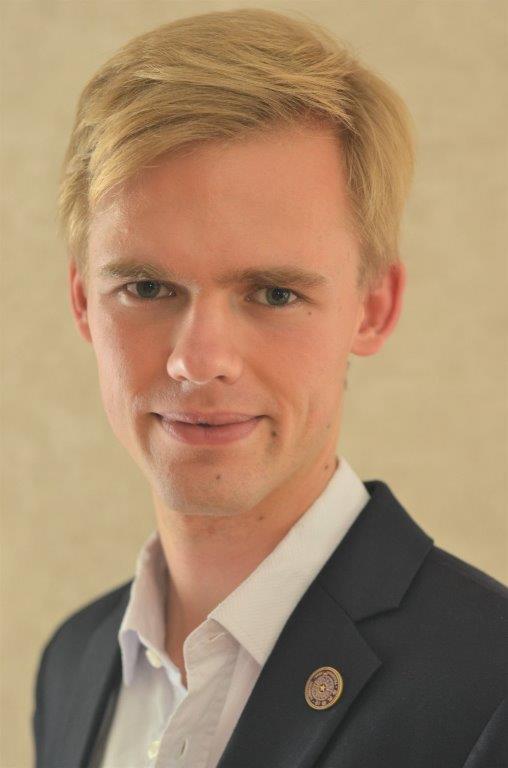}}]{Bernhard~Häfner}

    received a Master's degree in Electrical Engineering and Information
    Technology from Technical University of Munich (TUM), Germany in
    2018. In the same year, he joined BMW Group and the Chair of Connected
    Mobility at TUM as a PhD candidate. His main research interests are
    cooperative maneuvers among automated vehicles, from enabling
    technologies to application-level protocols.

\end{IEEEbiography}

\begin{IEEEbiography}
    [{\includegraphics[width=1in,height=1.25in,clip,keepaspectratio]{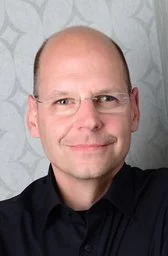}}]{Jörg~Ott}

    holds the Chair of Connected Mobility at Technical University of Munich in
    the Faculty of Informatics since August 2015.  He holds a PhD (1997) and a
    diploma in Computer Science (1991) from TU Berlin and a diploma in
    Industrial Engineering from TFH Berlin (1995).  His research interests are
    in network architectures, (transport) protocols, and algorithms for
    connecting mobile nodes to the Internet and to each other.  He explores
    edge and in-network computing as well as decentralized services.  His
    research includes measuring and modeling human mobility and Internet usage
    as basis for design and evaluation.

\end{IEEEbiography}

\begin{IEEEbiography}
    [{\includegraphics[width=1in,height=1.25in,clip,keepaspectratio]{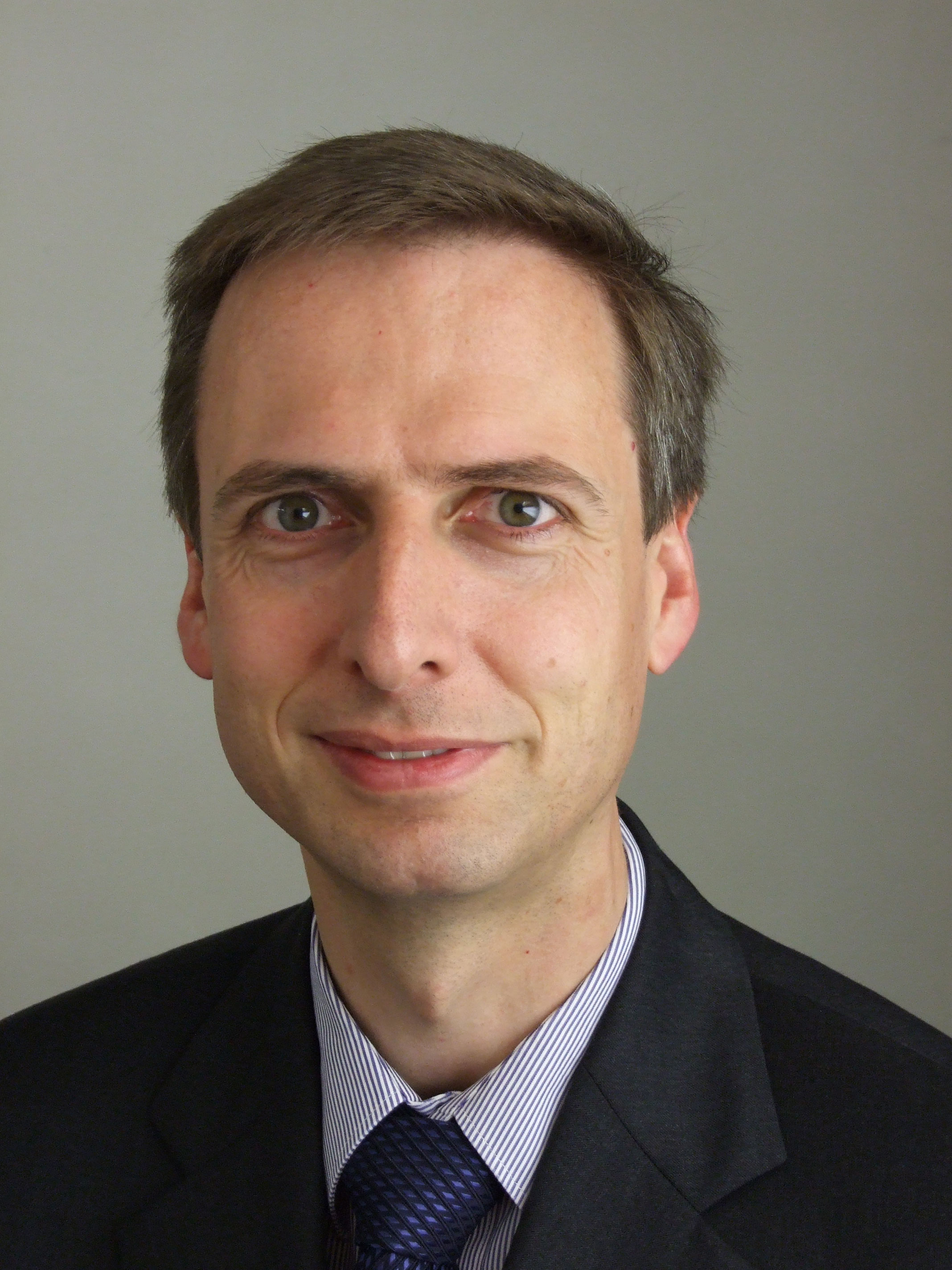}}]{Georg~A.~Schmitt}
    received his diploma and Ph.D. degrees in Physics from Technical
    University of Munich in 1992 and 1996, respectively.

    In 2012, he joined the BMW Group. He has worked as team lead in the
    area of charging infrastructure and standardization of charging interfaces,
    participating in and leading several German and EU INEA/CEF funded
    projects. Currently, he works as project lead responsible for vehicle
    connectivity technologies. He leads the activities on 5G and beyond
    cellular connectivity and ITS with focus on V2X solutions for automated and
    autonomous mobility.
    
    Dr. Schmitt actively takes part in standardization within the ISO as well
    as industry associations like 5GAA or VDA.

\end{IEEEbiography}

\vfill

\end{document}